\documentclass[aps,twocolumn,showpacs]{revtex4}
\usepackage{amsfonts}
\usepackage{amsthm}
\usepackage{amsmath}
\usepackage{amsfonts}
\usepackage{latexsym}
\usepackage{amssymb}
\usepackage{xcolor}
\usepackage[ansinew]{inputenc}
\usepackage{graphicx,fancyhdr}
\usepackage{lastpage}
\usepackage{epstopdf}
\usepackage{graphics}
\usepackage{latexsym}
\usepackage{amsmath}
\usepackage{dsfont}
\def\beqn{\begin{eqnarray}}
\def\eeqn{\end{eqnarray}}

\def\hw{\hbar \omega}
\def\hw4{ \frac {\hbar \omega}{4}}

\begin{document}
\title{Dynamics of finite dimensional non-hermitian systems with indefinite metric}
\vskip2cm
\author{Romina Ram\'\i rez $^{a)}$ \footnote{e-mail: romina@mate.unlp.edu.ar}}
\author{Marta Reboiro $^{b)}$ \footnote{e-mail: reboiro@fisica.unlp.edu.ar}}
\affiliation{{\small\it $^{a)}$Department of Mathematics, University of La Plata.}
{\small \it La Plata, Argentina}}
\affiliation{{\small\it $^{b)}$IFLP, CONICET-Department of Physics, University of La Plata.}
{\small \it La Plata, Argentina}}
\date{\today}

\begin{abstract}
We discuss the time evolution of  physical finite dimensional systems which are modelled by non-hermitian Hamiltonians.
We address both general non-hermitian Hamiltonians and pseudo-hermitian ones. We apply the theory of Krein Spaces to construct metric operators and well-defined inner products. As an application, we study the stationary behavior of dissipative One Axis Twisting Hamiltonians. We discuss the effect of decoherence under different coupling schemes.
\end{abstract}


\maketitle

key words: non-hermitian Hamiltonians, indefinite metric, Krein Spaces, Decoherence and Spin Squeezing.

\section{Introduction}

Recent years have seen a growing interest in the study of non-hermitian Hamiltonians, particularly in relation to open quantum systems \cite{rotter,qther,disi,disib,rotter2,savannah,mikhail,mikhail1,gadella,open1,open2,longhi2}. Among these types of Hamiltonians, pseudo-hermitian operators play a central role. The formal beginning of this subject was due to Bender and Boettcher \cite{bender} in 1998. The Authors of \cite{bender} have proposed the study of the celebrated Hamiltonian  $H_{1} = p^{2} + x^{2}(ix)$, which has real spectrum and is not self-adjoint. The more relevant characteristic of this Hamiltonian, which belongs to the parametric family $H_{\varepsilon >0} = p^{2} + x^{2}(ix)^{\varepsilon}$, and of others Hamiltonians that were studied later \cite{Bagarello book,mikhail}, is that they are invariant respect to Parity-Time Reversal ($\mathcal{PT}$) symmetry. These types of Hamiltonians are a particular case of pseudo-hermitian operators \cite{quasiher}. They have proven to be very useful in the understanding of physical problems with manifiest $\mathcal{PT}$
symmetry, i.e. microwave cavities \cite{8}, atomic diffusion \cite{12}, electronic circuits \cite{15}, optical waveguide arrays \cite{joglekar}, quantum critical phenomena \cite{ueda,mikhail,mikhail1}.

In the study of a parametric family of non-hermitian Hamiltonians, it is usual to observe regions with different symmetry. These zones are determined by the properties of the spectrum. In the search of eigenvalues, an exceptional point occurs when the coalescence of two or more eigenvalues is accompanied by the coalescence of the corresponding eigenvectors. In finite dimension, exceptional points take place when the diagonalization of a Hamiltonian $H$ breaks down, so that it can only be reduced to Jordan block form \cite{H,FF,kazuki,rotter1,brody,mikhail1,eva}. The existence of exceptional points has been visualized in various laboratory experiments \cite{lab1, lab2, lab3,longhi,pe1,pe2}.

In most studies, the focus is on the region of unbroken symmetry \cite{disi,benderrep,bender89,nososc16,mikhail}. Recently, the Authors of \cite{bender5} have studied the region with broken symmetry for the family of Hamiltonians $ H_{\varepsilon}$. Their findings have clarified the existence of divergences in different perturbative developments \cite{benderrep, FF3, benderdiv}.

The formalization of the time evolution of the observables of physical systems, which are described by non-hermitian Hamiltonians, is related to the introduction of well-defined inner products \cite{scholtz, benderrep,most2010,spiegel}. The literature devoted to time evolution of physical systems, which are modelled by non-hermitian Hamiltonians, is predominately focussed  in the determination of the survival probability of a particular component of the initial state as it evolves in time. The time evolution of physical observables have been mostly addressed by means of perturbative expansions \cite{per2,per1} or by phenomenological approaches, i.e master equation for the density matrix \cite{master}.

In this work we propose a formalism to study the time evolution \cite{rotter,faisal,feschbach,fringt,savannah} of a given initial state, in the presence of an interaction modelled by a finite dimensional non-hermitian Hamiltonian.
%We aim to study the mean value of the physical observables of the system as a function of time.
In the broadest sense, the Hamiltonian of an open
quantum system \cite{rotter} consists of a first order interaction
term describing a localized system with
discrete states and a second-order term caused by the
interaction of the discrete states with an external environment.
It can be distinguished two very different
cases of coupling with the environment. In the first case the environment consists of a continuum of scattering
wave functions which can mediate the escape of particles from the localized system, i.e. unstable states in a nuclei \cite{feschbach,rotter}.
In the other case, the environment is provided by the states of a macroscopic reservoir, and the strength of coupling depends on the overlap between states of the localized system and states of the reservoir, i.e. the transport of
electrons through mesoscopic quantum dots \cite{ferry}, or the engineering of tight-binding quantum networks \cite{longhi2}.
A widely used approach to study open quantum systems is the Feshbach projection
operator formalism \cite{rotter,faisal,feschbach}. In
this approach, the system under study is divided in two subspaces, the subspace corresponding
to the localized system, and the subspace
related to the environment. The
solution of problem in the whole function space (localized system embedded in a well-defined environment), which it is described by an hermitian Hamiltonian operator, can be represented in the interior of the localized
part of the system, after applying the Feshbach's formalism, by a set of eigenfunctions of an
effective non-hermitian Hamiltonian. The corresponding matrix elements describing the coupling
that develops between the different states of the localized system are typically
complex, consisting of real and imaginary parts, to account for the interaction with the external environment.
The reader is kindly referred to \cite{feschbach,rotter,faisal} for further details.

A non-hermitian model of  physical interest is the Hamiltonian that describes the interaction of a system of N collective spins interacting through a dissipative non-hermitian One Axes Twisting (OAT) interaction. It takes the form
\beqn
H & = &-\frac \omega 2+ H_{OAT}+ H_d, \label{hami}\\
H_{OAT} & = &  -\frac 12 \lambda ~ S_z^2, \nonumber \\
H_d& = &  + {\bf i }~ 2 \kappa ~ S_x. \nonumber
\eeqn
The components of the collective pseudo-spin operator of the system, ${\bf S}= \left( ~S_x,~S_y,~S_z \right)$, obey the cyclic commutation relations $\left[~S_i,~S_j\right]=~{\rm \bf i}~\epsilon_{ijk}~S_k$, where the suffixes $i,j,k$ stand for the components of the spin in three orthogonal directions and $\epsilon_{ijk}$ is the Levi-Civita symbol. The corresponding Hilbert space $\mathcal{H}_{S}$ has dimension $2S+1$.

In writing the collective Hamiltonian of Eq.(\ref{hami}), we assume that we have a system of $N$ collective spins interacting among themselves and with an external system. Instead of work with the Hamiltonian of the whole system, we model the effects of the interaction with the external system through the non-hermitian term $H_d$. Physically, it can be said that $H_d$ accounts for the effects of decoherence of the system due to the interactions with its environment \cite{Schlosshauer,zurek}.
The term $H_{OAT}$ is the OAT interaction introduced by Kitagawa and Ueda in \cite{kitagawa} to model the effect of Spin Squeezing.
Different systems can be modeled by Hamiltonians closely related
to one proposed in Eq.(\ref{hami}), i.e. a system of two-component atomic condensates \cite{condensates}, or  dissipative systems of solid-state spins in diamond \cite{nvnew,nosap17}.

%This parametric family of Hamiltonians is also interesting from the mathematical point of view. The spectrum of  $H$ can be classified according %to the values of the parameters $\omega, \lambda$ and $\kappa$.

From a mathematical point of view, the Hamiltonian of Eq.(\ref{hami}) can be taken as a parametric family of pseudo-hermitian Hamiltonians, which are invariant under ${\mathcal PT}$ symmetry \cite{bender89}.
The linear parity operator ${\mathcal P}$ performs spatial reflection, so that the position, the momentum and the spin transform as
${\bf r} \rightarrow - {\bf r}$,
${\bf p} \rightarrow - {\bf p}$, and
${\bf S} \rightarrow   {\bf S}$, respectively. Whereas,
time-reversal operation can be represented by an anti-unitary
operator ${\mathcal T}=U~ K$, being $U$  an unitary operator and $K$ the complex conjugation operator \cite{bhor-mottelson,trsu2}. Under time reversal, we have
${\bf r} \rightarrow  {\bf r} $,
${\bf p} \rightarrow  {\bf p} $,
${\bf S} \rightarrow -{\bf S} $ and
${\rm \bf i} \rightarrow -{\rm \bf i}$.
For the $su(2)$ spin algebra, time-reversal operator can be realized by ${\mathcal T}={\rm e}^{{\bf i} \pi S_y}~ K$ \cite{trsu2}.

The behavior of the spectrum of $H$ is a consequence of the invariance of $H$ under ${\mathcal PT}$ symmetry, $H={\mathcal T}{\mathcal P}H{\mathcal P}^{-1}{\mathcal T}^{-1}$. Depending on the values of the family parameters, ($\omega$, $\kappa/\lambda$, N), the spectrum of Hamiltonian of Eq.(\ref{hami}) is real, it means that ${\mathcal PT}$ is not spontaneously broken, i.e. the eigenfunctions
of H are simultaneously eigenfunctions of ${\mathcal PT}$. For other values of ($\omega$, $\kappa/\lambda$, N), ${\mathcal PT}$ symmetry is spontaneously broken, the eigenfunctions of $H$ are no longer eigenstates of ${\mathcal PT}$, and the spectrum of $H$ contains
complex-conjugate pairs. At fix number of spins and for some particular values of the ratio $\kappa/\lambda$, the so called exceptional points, the coalescence of some eigenvalues are present.

The work is organized as follows. The details of the general formalism are presented in Section \ref{formalism}. We construct metric operators and its corresponding inner products, in order to evaluate mean value of the observables as a function of time. We discuss each of the possible scenarios, i.e. Hamiltonians with real eigenvalues, with complex-conjugate pair eigenvalues, existence of exceptional points and Hamiltonians with general complex eigenvalues.
The results of the calculations, applied to the Hamiltonian of Eq.(\ref{hami}), are presented and discussed in Section \ref{results}.  Our conclusions are drawn in Section \ref{conclusions}.

\section{Formalism}\label{formalism}

In what follows, we shall present the formalism to describe the dynamics of a general non-hermitian Hamiltonian $H$ acting in a finite dimensional $\mathcal{H}$ Hilbert space.
Our aim is to compute the mean value of a physical observable, when a given initial state evolves in time, under the action of a non-hermitian Hamiltonian. We shall represent the physical observable by the linear hermitian operator $\widehat{o}$. To calculate the expectation value of $\widehat{o}$, we work with the basis, $\mathcal{A}_{H}$, formed by the eigenstates or generalized eigenstates of $H$ and we look for a metric operator $\mathcal{S}$, i.e. an operator which is self-adjoint and positive definite, in order to construct an inner product $\langle . | . \rangle_{\mathcal S}$.
The Hilbert space ${\mathcal H}$ equipped with the inner product $\langle  . | . \rangle_{\mathcal S}$ is the new physical linear space ${\mathcal H}_{\mathcal S}=(\mathcal{H}, \langle .| . \rangle_{\mathcal S})$. Over this Hilbert space, we calculate well-defined expectation values.

A particular case of non-hermitian operators are the so called pseudo-hermitian Hamiltonians. We say that an operator $H$ in a Hilbert space $\mathcal{H}$ is pseudo-hermitian (with respect to $S$) if $H$ is densely defined in $\mathcal{H}$ and there exists a bounded self-adjoint
operator $S$ with bounded inverse $S^{-1}$ such that
$H^{\dagger} = SHS^{-1}$. Any pseudo-hermitian operator is closed and its spectrum consists of real or complex-conjugate pair eigenvalues, that is $H$ and $H^\dagger$ are isospectral operators. In finite dimensional Hilbert space operator $S$ is always bounded, furthermore it fulfills the relation $S H = H^\dagger S$.

In dealing with a pseudo-hermitian Hamiltonian $H$, we shall assume that it is a particular element of a parametric family of Hamiltonians $ H_{\delta} $. This parameter or set of parameters, $\delta$, is in direct relation with the coupling constants of the physical problem under consideration.
In general, the properties of the spectrum of $H$ varies throughout the parameter space, i.e. real spectrum or spectrum which includes complex-conjugate pair eigenvalues. We shall call exceptional points to those values of $ \delta $ for which the Hamiltonian $H_{\delta}$, in finite dimension, is not diagonalizable.

%(see \ref{Appendix}).
%Let us consider the different possible scenarios, i.e. real eigenvalues, complex pair-conjugate eigenvalues, exceptional points and general complex %eigenvalues.

Let us briefly review the main properties associated to the spectrum of non-hermitian Hamiltonians \cite{faisal}. We can write the action of $H$ on an orthonormal basis $\mathcal{A}_k$ of $\mathcal{H}$. From the representation of $H$ in the basis $\mathcal{A}_k$, we obtain eigenfunctions of $H$, $\mathcal{A}_{H} = \{ |\widetilde{\varphi}_{j} \rangle \}_{ j=1...N_{max}}$, i.e
\begin{equation}
H | \widetilde{\varphi} _{j} \rangle=\widetilde{E}_{j} |\widetilde{\varphi} _{j} \rangle.
\label{avaH}
\end{equation}
In the same way, the set of eigenfunctions of $H^\dagger$, $\mathcal{A}_{H^{\dagger}}=\{ |\psi_{i} \rangle \}_{ i=1...N_{max}}$, satisfies
\begin{equation}
H^{\dagger} |\bar{\psi} _{j} \rangle= \bar{E}_{j} | \bar{\psi}_j \rangle,
\label{avaHd}
\end{equation}
When working with systems of infinite dimension, the sets $\mathcal{A}_{H}$ and $\mathcal{A}_{H^{\dagger}}$, not always form a basis \cite{bender,Bagarello book}. Nevertheless, in finite dimension it is straightforward to show that, if $H$ is diagonalizable, the sets $\mathcal{A}_{H^{\dagger}}$ and $\mathcal{A}_{H}$ form a bi-orthonormal set of $\mathcal{H}$ \cite{Sw}, i.e.
\begin{eqnarray}
\langle \bar{\psi}_{i}| \widetilde{\varphi}_{j}\rangle =  \delta_{ij},
\label{biorthonormal}
\end{eqnarray}
with
\beqn
\bar{E}_{j}=\widetilde{E}^*_{j}.
\label{eigenvalue}
\eeqn

In the next Sections, we shall construct the metric operator $ \mathcal{S}$ for the different classes of non-hermitian Hamiltonians. Particularly, in the case of pseudo-hermitian Hamiltonians with broken symmetry, we shall make use the formalism of Krein Spaces \cite{Krein69}.

\subsubsection{Case I. Pseudo-hermitian diagonalizable Hamiltonian: Real spectrum}\label{case1}

Let $H$ a pseudo-hermitian diagonalizable Hamiltonian with  real spectrum.
In this case, we can define a symmetry operator ${\mathcal S}_{\psi}$ so that ${\mathcal S}_{\psi} | \widetilde{\phi}_j \rangle= |\bar{\psi }_j\rangle$. In terms of the eigenvectors of $H^\dagger$ it is given by

\begin{equation}
{\mathcal S}_{\psi}= \sum_{j=1}^{N_{max}}~|\bar{\psi}_j \rangle \langle \bar{\psi}_j|,
\label{sphi}
\end{equation}
and it obeys ${\mathcal S}_{\psi} H= H^{\dagger}{\mathcal S}_{\psi}$. The symmetry operator $S_{\psi}$ is self-adjoint and positive, so that we can define an inner product on $\mathcal{H}$ by
\beqn
\langle f | g \rangle_{{\mathcal S}_{\psi}}= \langle f | {{\mathcal S}_{\psi}} g \rangle.
\label{inn1}
\eeqn

The Hilbert space $\mathcal{H}$ equipped with the inner product $\langle  . | . \rangle_{{\mathcal S}_{\psi}}$ is the new physical Hilbert space $\mathcal{H}_{{\mathcal S}_{\psi}}:=(\mathcal{H}, \langle .| . \rangle_{{\mathcal S}_{\psi}})$ where the expectations values for the time evolution can be formally calculated.

\subsubsection{Case II. Pseudo-hermitian diagonalizable Hamiltonian: Non-degenerate complex-conjugate pair spectrum}\label{case2}

If the spectrum of $H$ includes non-degenerate complex-conjugate pair eigenvalues, the operator ${\mathcal S}_{\psi}$ of Eq.(\ref{sphi}) is not longer a metric operator and ${\mathcal S}_{\psi} H \neq H^{\dagger}{\mathcal S}_\psi$.

The self-adjoint symmetry operator, which enables us to recovery the property ${\mathcal S}H= H^{\dagger}{\mathcal S}$, can be written as

\beqn
{\mathcal S}
& = & \sum_{j \le i}^{N_{max}}
~ \delta( \bar{E}_j-\bar{E}^*_i) ~ \left( \alpha_{j}|  \bar{\psi}_{j} \rangle  \langle \bar{\psi}_{i}|+
~ \alpha^*_{j}|  \bar{\psi}_{i} \rangle  \langle \bar{\psi}_{j}| \right ), \nonumber \\
\eeqn
with $\alpha \in {\mathds C}$, and ${\rm Im}(\alpha) \neq 0$. However, ${\mathcal S}$ is not positive definite. Thus, the inner product $[x,y]=(x,{\mathcal S}y)$ is indefinite. This problem can be avoided by considering the decomposition of ${\mathcal S}$ in a the positive and a negative part. This decomposition is framed within the theory of Krein spaces \cite{Krein69}.

As ${\mathcal S}$ is a self-adjoint and diagonalizable operator, its eigenvalues can be classified according to their sign, semipositive $\lambda_{+i} $ or negative $\lambda_{-j}$. We can decompose $\mathcal{H}$ as a direct sum $\mathcal{H}^{+} \oplus \mathcal{H}^{-}$, where $\mathcal{H}^{+}$ is spanned by eigenfunctions corresponding to $\{\lambda_{+i}\}$ and $\mathcal{H}^{-}$ is spanned by eigenfunctions corresponding to $\{\lambda_{-j}\}$, respectively. Then ${\mathcal S}=P{\mathcal D}P^{-1}$, where ${\mathcal D}$ is the diagonal matrix containing the eigenvalues $\{\lambda_{+1},.....\lambda_{+M}, \lambda_{-1},...\lambda_{-N} \}$ with $N+M=N_{max}$. Matrix $P$ contains respective eigenvectors.

With this, ${\mathcal D}={\mathcal D}_+ + {\mathcal D}_-$, being

\beqn
{\mathcal D}_+  =\left(
\begin{array}{ll}
{\bf d_{+}} & {\bf 0} \\
{\bf 0} & {\bf 0}
\end{array}
\right), ~~~~
{\mathcal D}_-  =\left(
\begin{array}{ll}
{\bf 0} & {\bf 0} \\
{\bf 0} & {\bf d_-}
\end{array}
\right),
\eeqn
the positive and negative parts of ${\mathcal D}$.

We shall define, in $\mathcal{H}$, the operators ${\mathcal S}_{+}=P{\mathcal D}_{+}P^{-1}$ and ${\mathcal S}_{-}=P {\mathcal D}_{-}P^{-1}$.
Operators  ${\mathcal S}_{\pm}$ are self-adjoint, being ${\mathcal S}_+$ positive definite and ${\mathcal S}_-$ negative definite.

Then, we can introduce the metric operator ${\mathcal S}_K={\mathcal S}_{+}-{\mathcal S}_{-}$  in $\mathcal{H}$. The inner product $\langle .|.\rangle_{{\mathcal S}_K}$ in  $\mathcal{H}$ is defined as
\beqn
\langle x|y \rangle_{{\mathcal S}_K}=
\langle~ x~|~{\mathcal S}_K ~ y \rangle =
\langle~ x^{+}~| ~ {\mathcal S}_+~ y^{+}\rangle~ -~ \langle x^{-}~ | ~{\mathcal S}_{-}y^{-}~\rangle, \nonumber \\
\eeqn
for all $x, y \in \mathcal{H}$, being $x^{\pm}$ and $y^{\pm}$ its components in the canonical decomposition.

As ${\mathcal S}_K$ is a metric, it can be written as ${\mathcal S}_K=\Upsilon^\dagger_K \Upsilon_K$,
with $\Upsilon_K= ({{\mathcal S}_+}^{1/2}+{{\mathcal S}_-}^{1/2})$,
being ${{\mathcal S}_\pm}^{\pm 1/2}= P^{-1} {{\mathcal D}_\pm}^{\pm 1/2} P$, with
\beqn
{{\mathcal D}_+}^{\pm 1/2}  =\left(
\begin{array}{ll}
{\bf {d_+}}^{\pm 1/2} & {\bf 0} \\
{\bf 0} & {\bf 0}
\end{array}
\right), ~~~~
{\bf{\mathcal D}_-}^{\pm 1/2}  =\left(
\begin{array}{ll}
{\bf 0} & {\bf 0} \\
{\bf 0} & {\bf {d_-}}^{\pm 1/2}
\end{array}
\right), \nonumber \\
\eeqn
and where
$({\bf {d_+}}^{\pm 1/2})_{ij}=\delta_{i j} ({\lambda_{+i}})^{\pm 1/2}$, while the entries of ${\bf d_-}$ are $({\bf d_-}^{\pm 1/2})_{ij}=\delta_{i j}({\lambda_{-i}})^{\pm 1/2}$. We shall work in the basis of ${\mathcal S}$, ${\mathcal A}_S$, which does not coincide with ${\mathcal A}_H$.

To preserve the mean value of an observable $\widehat{o}$, we take the new physical space $(\mathcal{H}, \langle .|. \rangle_{S_{K}})$, with
\beqn
S_{K}&=&PDP^{-1}, \nonumber \\
D&=&{\mathcal D}_{+} -{\mathcal D}_{-},
\label{diag}
\eeqn
clearly, $D$ is a real and positive definite diagonal matrix.

As it has been discussed in \cite{fring0}, to fix the metric uniquely, such there is no ambiguities in the interpretation of physical observables, we shall assume that matrix representation of the  hermitian $\widehat{o}$ on the basis ${\mathcal A}_S$, transforms as

\beqn
[\widehat{o}]_{{\mathcal A}_S} \rightarrow O=D^{-1/2}~ [\widehat{o}]_{{\mathcal A}_S} ~ D^{1/2},
\label{changebasis}
\eeqn
and that the coordinates of the vectors transform as
\beqn
[{\bf f}]_{{\mathcal A}_S} \rightarrow {\bf {\rm F}}= D^{-1/2} [{\bf f}]_{{\mathcal A}_S}.
\eeqn
So that,
\beqn
\langle f | \widehat{o} | g \rangle_{{\mathcal S}}& = & {\bf {\rm F}}^\dagger~D~O~{\bf {\rm G}}, \nonumber \\
& = & \sum_{\alpha \beta}~ {\widetilde{f}}_\alpha^*~ { \left ( [\widehat{o}]_{{\mathcal A}_H} \right ) }_{\alpha \beta}~{\widetilde g}_\beta, \nonumber \\
& = & \sum_{k l}~f_k^*~ {\left([\widehat{o}]_{{\mathcal A}_k} \right)}_{k l}~ g_l. \nonumber \\
\label{finners}
\eeqn
where ${\bf {\rm G}}= D^{-1/2} [{\bf g}]_{{\mathcal A}_S}$.

\subsubsection{Case III. Pseudo-hermitian Hamiltonian: Exceptional Points} \label{case3}

In finite dimension, when $H$ is not longer diagonalizable, we have to make use of the Jordan Decomposition. In this case, $H$ can be written as  $H=\widetilde{P}{\mathcal J}\widetilde{P}^{-1}$. Generalized eigenvectors constitute the columns of the matrix $\widetilde{P}$ and form a basis of $\mathcal{H}$. In the same way, $H^\dagger=\bar{P} {\mathcal J}^{\dagger} \bar{P}^{-1}$, with
$\bar{P}= ({\widetilde P}^{\dagger})^{-1}$.
Let $|\bar{\psi}_{k} \rangle$ being the $k^{th}$ column of $\bar{P}$.
Vectors $\{ | \bar{\psi}_{k} \rangle \}_{1 \leq k \leq N_{max}}$ form a non-orthonormal basis of $\mathcal{H}^\dagger$. As $\bar{P}^{-1}\bar{P}=I$, the set $\{ |\bar{v}_{k} \rangle \}_{1 \leq k \leq N_{max}}$, where $| \bar{v}_{k} \rangle$ is the $k^{th}$ column of $(\bar{P}^{-1})^{T}$, forms a basis of $\mathcal{H}^\dagger$, which is bi-orthonormal to
$\{|\bar{\psi}_{k} \rangle \}_{1 \leq k \leq N_{max}}$, i.e $ \langle \bar{v}_{k} | \bar{\psi}_{j} \rangle = \delta_{k, j}$.

Let us construct a new self-adjoint symmetry operator as

\beqn
{\mathcal S}_{J}= \sum_{j \le i}^{N_{max}}
~ \delta( \bar{E}_j-\bar{E}^*_i)~ \left ( \alpha_{j}|  \bar{\psi}_{j} \rangle  \langle \bar{v}_{i}|+
\alpha^*_{j}|  \bar{\psi}_{i} \rangle  \langle \bar{v}_{j}| \right). \nonumber \\
\label{sympe}
\eeqn
It is straightforward to prove that ${\mathcal S}_{J} H=H^{\dagger}{\mathcal S}_{J}$.
As before, ${\mathcal S}_{J}$ is a non-positive definite operator, so $[f,g]_{{\mathcal S}_J}=\langle f | {\mathcal S}_J g \rangle $ is an indefinite inner product for $\mathcal{H}$.

As $\mathcal{S}_J$ is a self-adjoint operator, we can follow the steps of Section \ref{case2}. After diagonalization of $\mathcal{S}_J$, it reads
$\mathcal{S}_J=R {\mathcal D}_J R^{-1}$. As before, $\mathcal{D}_J=\mathcal{D}_{J+}+\mathcal{D}_{J-}$ with

\beqn
{\mathcal D}_{J+}  =\left(
\begin{array}{ll}
{\bf d_{J+}} & {\bf 0} \\
{\bf 0} & {\bf 0}
\end{array}
\right), ~~~~
{\mathcal D}_{J-}  =\left(
\begin{array}{ll}
{\bf 0} & {\bf 0} \\
{\bf 0} & {\bf d_{J-}}
\end{array}
\right),
\eeqn
Again, we shall define in $\mathcal{H}$ the operators ${\mathcal S}_{\pm J}=R{\mathcal D}_{J \pm} R^{-1}$  and decompose $\mathcal{H}=\mathcal{H}^+ \oplus \mathcal{H}^{-}$. Both operators are self-adjoint, being ${\mathcal S}_{+J}$ positive definite and $S_{-J}$ negative definite.

At this point, we are in condition to  introduce the metric operator ${\mathcal S}_{KJ}={\mathcal S}_{+J}-{\mathcal S}_{-J}$, which is self-adjoint and positive definite. Also, $ \mathcal{S}_{KJ}={\Upsilon^\dagger_{KJ}} \Upsilon_{KJ}$. Consequently, we shall  define the inner product $\langle .|.\rangle_{{\mathcal S}_{KJ}}$ in  $\mathcal{H}$ as

\beqn
\langle f | g \rangle_{{\mathcal S}_{KJ}}= \langle f | {{\mathcal S}}_{KJ}~ g \rangle.
\label{finnerpe}
\eeqn

As before, we preserve the mean value of an observable $\widehat{o}$, by following the steps that we have presented in Eqs. (\ref{changebasis}-\ref{finners}), with $[\mathcal S]_{{\mathcal A}_S}=\mathcal{D}={\mathcal D}_{J+} - {\mathcal D}_{J-}$.

\subsubsection{Case IV. Non-pseudo-hermitian Hamiltonian}\label{case4}

If the spectrum of $H$ contains complex eigenvalues, which are not complex-conjugate pairs, $H$ and $H^\dagger$ are not isospectral Hamiltonians, the eigenvalues of $H^\dagger$ (${\bar E}_j$) are related to the eigenvalues of $H$ (${\tilde E}_j$) by (Eq.(\ref{eigenvalue})). In this context, we can define a new inner
product by introducing the operator

\beqn
{\mathcal S}_g = \sum_{j=1}^{N_{max}} ~ |  \bar{\psi}_{j} \rangle  \langle \bar{\psi}_{j}|.
\label{opSAg}
\eeqn
As $H$ is no longer a pseudo-hermitian operator, it results that $\mathcal{S}_g H \neq H^\dagger \mathcal{S}_g$.

The operator ${\mathcal S}_g$ of Eq.(\ref{opSAg}) is self-adjoint and positive, so that we can define an operator $\Upsilon_g$ such that $\mathcal{S}_g={\Upsilon^\dagger_g} \Upsilon_g$.  We are in condition to introduce an inner product on $\mathcal{H}$ of the form

\beqn
\langle f | g \rangle_{{\mathcal S}_g}= \langle f | {{\mathcal S}_{g}} g \rangle.
\label{inn1}
\eeqn
The Hilbert space $\mathcal{H}$ equipped with the inner product $\langle  . | . \rangle_{{\mathcal S}_{g}}$ is the new physical Hilbert space $\mathcal{H}_{{\mathcal S}_{g}}=(\mathcal{H}, \langle .| . \rangle_{{\mathcal S}_{g}})$.

As in Case I, Section \ref{case1}, the basis ${\mathcal A}_S$ coincides with ${\mathcal A}_H$ and $\mathcal{D}$ is the identity matrix.

We can summarize the previous results as follows. We have constructed, depending on the characteristics of the spectrum of $H$, a self-adjoint positive definite operator, $\mathcal{S}$ that allows to define an inner product. In this way, the mean values of physical observables can be properly  computed.

\subsection{Time Evolution}\label{timeevol}

We shall construct the time evolution of  a general initial state, $|I \rangle$. In the basis ${\mathcal A}_k$, it reads

\beqn
| I \rangle= \sum_k ~ c_k ~ | k \rangle.
\label{ini00}
\eeqn
In terms of the basis formed by the eigenvectors of $H$ the initial state can be written as

\beqn
| I \rangle & = & \sum_\alpha ~ \widetilde{c}_\alpha ~ |\widetilde{\phi}_ \alpha \rangle, \nonumber \\
\widetilde{c}_\alpha & = & \sum_k~ (\Upsilon^{-1})_{\alpha k}~ c_k,
\label{ini0}
\eeqn
with $\Upsilon$ the transformation matrix from basis the ${\mathcal A}_k$ to the basis ${\mathcal{A}}_H $.
We shall assume that the initial state is normalized,
that is $\langle {I} | {I} \rangle=1$. The initial state of Eq.(\ref{ini0}) evolves in time as

\beqn
| I(t) \rangle & = & {\rm e}^{- i H t} | I \rangle, \nonumber \\
  & = & \sum_\alpha ~ {{\widetilde c}_\alpha(t)} ~ |\widetilde{\phi}_\alpha \rangle.
% \nonumber \\
%  & = & \sum_l ~ c_l(t) ~ |l \rangle, \nonumber \\
%             c_l(t)   &=& \sum_\alpha~ \Upsilon_{l \alpha}~\widetilde{c}_\alpha(t).
\label{init}
\eeqn
If $H$ can be diagonalized, $\widetilde{c}_\alpha(t)$ is given by $\widetilde{c}_\alpha(t)={\rm e}^{- i \widetilde{E}_\alpha t}~\widetilde{c}_\alpha$.
In the case of exceptional points, the Hamiltonian $H$ can be decomposed in terms of the Jordan matrix $J$, as $H=\Upsilon {\rm e}^{- i J t} \Upsilon^{-1}$. Correspondingly, the form of the coordinates ${{\widetilde c}_\alpha(t)}$ will depend upon the particular Hamiltonian under consideration.

In terms of the eigenvectors of the symmetry operator $\mathcal{S}$, the initial state reads

\beqn
|I(t) \rangle & = &
\sum_\beta~
{\overset{\approx}{c}}_\beta(t)~|{\overset{\approx}{\phi}}_\beta \rangle,\nonumber \\
{\overset{\approx}{c}}_\beta(t)& = & \sum_{\alpha}~
(\Upsilon'^{-1})_{\beta \alpha}~  \widetilde{c}_\alpha(t),
\label{inits}
\eeqn
with $\Upsilon'$ being the transformation matrix from  the  basis  ${\mathcal A}_H$ to the basis ${\mathcal A}_S$.

At this point, we are in condition of evaluating
the mean value of an operator $\widehat{o}$ as a function of time as
\beqn
\langle \widehat{o}(t) \rangle & = & {\langle I (t)| \widehat{o}|I(t) \rangle}_{\mathcal{S}} \nonumber \\
& = & \sum_{\alpha \beta}~ {\overset{\approx}{c}}_\alpha(t) {\overset{\approx}{c}}^*_{\beta}(t)~
\left \langle {\overset{\approx}{\phi}}_\beta  \mid \widehat{o} \mid {\overset{\approx}{\phi}}_\alpha \right \rangle_{\mathcal{S}}.
\eeqn
As it has been stated before, in order to evaluate $\left \langle {\overset{\approx}{\phi}}_\beta | \widehat{o} |{\overset{\approx}{\phi}}_\alpha \right \rangle_{\mathcal{S}}$ we follow the prescription given in Eqs.(\ref{changebasis}-\ref{finners}).

\section{Physical Applications}\label{results}

Let us apply the previous results to the study of the time evolution of a given initial state, under the non-hermitian Hamiltonian that we have introduced in Eq.({\ref{hami}}).

We shall assume that the initial state is prepared as a coherent spin-state (CSS) given by

\begin{eqnarray}
|I (\theta_0,\phi_0)\rangle= {\cal N} \sum_{k=0}^{2 S}~ z(\theta_0,\phi_0)^k \left (\begin{array}{c} 2 S \\ k \end{array} \right)^{1/2} |k\rangle , \label{istate}
\end{eqnarray}
with $ z(\theta_0,\phi_0) ={\rm e}^{-i \phi_0} \tan(\theta_0/2)$. The angles $(\theta_0,\phi_0)$ define the direction
$\vec{n}_{0}=(\sin{\theta_0} \cos{\phi_0},\sin{\theta_0}
\sin{\phi_0},\cos{\theta_0})$, such that $\vec{S} \cdot \vec{n}_0 |I\rangle=-S | I\rangle$ \cite{hecht}.

An observable of interest related to the Hamiltonian of Eq.(\ref{hami}) is the spin squeezing parameter. Spin-squeezed-states are quantum-correlated states  with reduced fluctuations in one of the components of the total spin. Following the work of Kitagawa and Ueda \cite{kitagawa}, we shall
define a set of orthogonal axes $\{ {\bf n_{x'}}, {\bf n_{y'}},
{\bf n_{z'}} \}$, such that ${\bf n_{z'}}$ is the unitary vector pointing along the direction
of the total spin $<{\bf S }>$ .  We shall fix the direction ${\bf n_{x'}}$ looking for
the minimum value of $\Delta^2S_{x'}$, so that the Heisenberg  Uncertainty Relation reads

\beqn
%\Delta^2 S_{y'} ~\Delta^2 S_{x'}~&\ge& ~ \frac 14 |<S_{z'}>|^2\nnu
\Delta^2 S_{y'} ~\Delta^2 S_{x'}~& \ge & ~\frac 14 |<S_{z'}>|^2.
\label{hur}
\eeqn
Consequently, the squeezing parameters \cite{kitagawa} are defined as

\begin{eqnarray}
\zeta^2_{x'}  =  \frac {2 \Delta^2 S_{x'} }{|<S_{z'}>|},~
\zeta^2_{y'}  =  \frac {2 \Delta^2 S_{y'} }{|<S_{z'}>|}.
\label{sqx}
\end{eqnarray}
The state is squeezed in the $x'$-direction if $\zeta^2_{x'}<1$ and $\zeta^2_{y'}>1$.
If the minimum value of the Heisenberg Uncertainty Relation, Eq.(\ref{hur}), is achieved and $\zeta^2_{x'}<1$, the state is
called Intelligent Spin State \cite{iss1,iss2,iss6,nosiss}.

%\subsubsection{Pseudo-hermitian OAT Hamiltonian.}
In Figure 1, we show the results concerning the number of real eigenvalues of the Hamiltonian $H$ of Eq.(\ref{hami}), as a function of the ratio $\kappa/\lambda$, for systems with different number of particles, $N= 2 S$. Systems with an even number of particles always have, at least, one real eigenvalue, due to the fact that the space dimension is $2 S+1$. On the other hand, when the ratio $\kappa/\lambda$ is increased, systems with odd number of particles have not real eigenvalues. In what follows we shall describe the time evolution of systems with $N=4$ and $N=10$ particles, the corresponding points have been marked with crosses in the Figure.

\begin{figure}[h!]
\includegraphics[width=6cm]{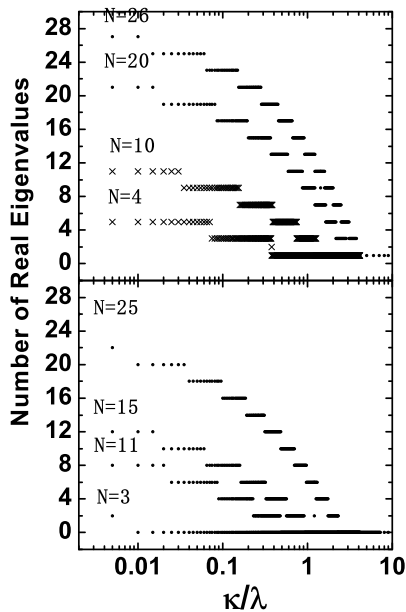}
\caption{Number of real eigenvalues of the $H$ of Eq.(\ref{hami}) as a function of the ratio $\kappa/\lambda$, for systems with different number of particles, $N= 2 S$.} \label{fig:fig1}
\end{figure}

In Figure 2, we display the results obtained for the Squeezing Parameters of Eq.(\ref{sqx}) as a function of the time, in units of $[{\rm dB}]$. We have considered a system with $N=10$ particles. Panels (a) and (c) show the results obtained for the coupling constants ratio $\kappa/\lambda=0.01$, while panels (b) and (d) correspond to $\kappa/\lambda=1.5$. In panels (a) and (b), we show the results obtained for an initial CSS with $(\theta_0,\phi_0)= (\pi/4,0)$. For panels (c) and (d) we have taken $(\theta_0,\phi_0)= (\pi/8,0)$ .
It is clear from the Figure that the time evolution of the initial state in the region of real spectrum of $H$ is quite different to that obtained in the region with complex spectrum. For small values of the ratio $\kappa/\lambda$, the model shows series of revivals, even when the term responsible for the decoherence of the system is not switched off, i.e. $\kappa \neq 0$. If the ratio $\kappa/\lambda$ is greater than 1 (See Figure 1), the number of real eigenvalues is reduced to one, and the initial state evolves into a steady state which behaves as an Intelligent State. As complementary information, in Figure 3, the mean value of the collective spin components of the system are displayed. Figure 3 confirms the series of revivals of the physical observables in the region of real spectrum, and the effect of decoherence, that is the existence of a pointer state, in the region with complex-conjugate pair eigenvalues.

\begin{figure}[h!]
\includegraphics[width=6cm]{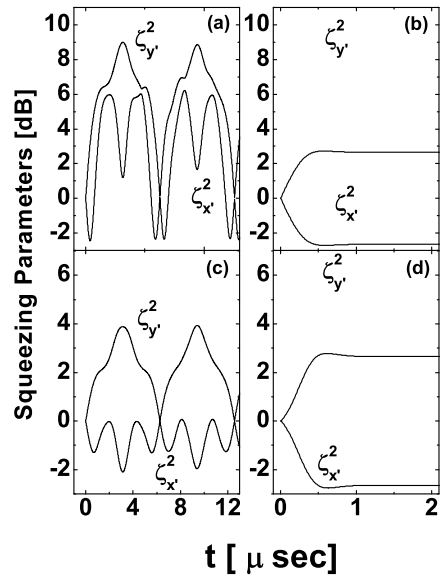}
\caption{Squeezing Parameters of Eq.(\ref{sqx}) as a function of the time, in units of $[{\rm dB}]$. Panels (a) and (c) show the results obtained for the coupling constants ratio $\kappa/\lambda=0.01$, while panels (b) and (d) correspond to $\kappa/\lambda=1.5$, for a sytem with $N=10$. In panels (a) and (b), we show the results obtained for an initial CSS with $(\theta_0,\phi_0)= (\pi/4,0)$. For panels (c) and (d) we have taken $(\theta_0,\phi_0)= (\pi/8,0)$ .} \label{fig:fig2}
\end{figure}

\begin{figure}[h!]
\includegraphics[width=6cm]{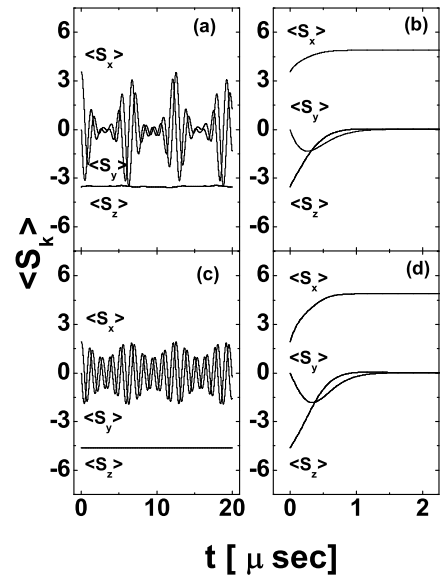}
\caption{Mean Values of the components of the spin, $\langle S_k \rangle$, as a function of the time. Panels (a) and (c) show the results obtained for the coupling constants ratio $\kappa/\lambda=0.01$, while panels (b) and (d) correspond to $\kappa/\lambda=1.5$. In panels (a) and (b), we show the results obtained for an initial CSS with $(\theta_0,\phi_0)= (\pi/4,0)$. For panels (c) and (d) we have taken $(\theta_0,\phi_0)= (\pi/8,0)$ .} \label{fig:fig3}
\end{figure}

The Hamiltonian of Eq.(\ref{hami}) has exceptional points \cite{H,FF,kazuki,rotter1,brody,mikhail1}. At these points, some eigenstates  coalesce and the Hamiltonian is not diagonalizable.

As an example, let us consider a system with $N=4$ particles.
In this case, there are two values of the coupling ratio $|\kappa/\lambda|$ at which the system has exceptional points. In Figure 4, we plot the eigenvalues of the system, as a function of the coupling ratio $\kappa/\lambda$. In panel (a) we show the behaviour of the real part of the eigenvalues, while in panels (b), we present the imaginary part of each eigenvalues. Clearly, exceptional points take place whenever $\kappa/\lambda=\pm 0.0739815$ or $\kappa/\lambda= \pm 0.375$. For values of $|\kappa/\lambda|<0.0739815$, the Hamiltonian has real eigenvalues, while for $|\kappa/\lambda| > 0.0739815$ complex-conjugate pair eigenvalues are present. At exceptional points the Hamiltonian can be decomposed as $H = P J P^{-1}$. Notice that, as $H=H^T$, $H^\dagger = {P}^* J^* \left(P^{-1}\right)^*$. In the present case the matrix $J$ takes the form

\beqn
J &= &
\left(
\begin{array}{ccccc}
 \widetilde{E}_1 & 1 & 0 & 0 & 0 \\
 0 & \widetilde{E}_1 & 0 & 0 & 0 \\
 0 & 0 & \widetilde{E}_3 & 0 & 0 \\
 0 & 0 & 0 & \widetilde{E}_4 & 0 \\
 0 & 0 & 0 & 0 & \widetilde{E}_5 \\
\end{array}
\right).
\eeqn

Let us consider the exceptional point $\kappa/\lambda=0.0739815$. At this point, the Hamiltonian has real eigenvalues, $\widetilde{E}_1=\widetilde{E}_2=2.24$, $\widetilde{E}_3=0.514$, $\widetilde{E}_4= 0.515$ and $\widetilde{E}_5= 1.99$. The symmetry operator, ${\mathcal{S}}_J$ of Eq.(\ref{sympe}), takes the form

\beqn
{\mathcal{S}}_J=\sum_{k=1}^5~| {\bar{\psi}}_k \rangle \langle{\bar v}_k |,
\nonumber \\
\eeqn
where $| {\bar{\psi}}_k \rangle$ are the columns of $P^*$, and $| {\bar v}_k \rangle$ are the files of $(P^{*})^{-1}$.
After diagonalization, ${\mathcal{S}}_J=R~ (\mathcal{D}_{J+} + \mathcal{D}_{J-}) ~R^{-1}$
Thus, the matrix representation of the metric operator $\mathcal{S}_{KJ}$, which we use to define the inner product of Eq.(\ref{finnerpe}), in the basis $\mathcal{A}_S$, is $[\mathcal{S}_{KJ}]_{{\mathcal A}_S}= \mathcal{D}_{J+} - \mathcal{D}_{J-}$.

At the other exceptional point of the Hamiltonian of Eq.(\ref{hami}), $\kappa/\lambda=0.375$, the Hamiltonian has three real eigenvalues, $\widetilde{E}_1=\widetilde{E}_2=1.25$, $\widetilde{E}_3=0.754$, and two complex eigenvalues, $\widetilde{E}_4= 2.12-1.34 {\bf i}$ and $\widetilde{E}_5= 2.12+1.34 {\bf i}$.
The symmetry operator, ${\mathcal{S}}_J$ of Eq.(\ref{sympe}), takes the form

\beqn
{\mathcal{S}}_J=\sum_{k=1}^3~| {\bar{\psi}}_k \rangle \langle{\bar v}_k |~+~{\bf i}~| {\bar{\psi}}_4 \rangle \langle {\bar v}_5 |-~{\bf i}~| {\bar{\psi}}_5 \rangle \langle {\bar v}_4 |,
\nonumber \\
\eeqn
where $| {\bar{\psi}}_k \rangle$ are the columns of $P^*$, and $| {\bar v}_k \rangle$ are the rows of $(P^{*})^{-1}$.
After diagonalization, ${\mathcal{S}}_J=R~ \mathcal{D}_J ~R^{-1}$, with $\mathcal{D}_J=\mathcal{D}_{J+} + \mathcal{D}_{J-}$, being $\mathcal{D}_{J+}$ and $\mathcal{D}_{J-}$, the matrices with positive and negative eigenvalues of ${\mathcal{S}}_J$ in the diagonal, respectively.

Then, the matrix representation of the metric operator $\mathcal{S}_{KJ}$, which it is used in the definition the inner product of Eq.(\ref{finnerpe}), in the basis $\mathcal{A}_S$, is $[\mathcal{S}_{KJ}]_{{\mathcal A}_S}={\rm D}=\mathcal{D}_{J+} - \mathcal{D}_{J-}$.

Concerning the time evolution of the initial state $|I \rangle$ of Eq.(\ref{ini00}), it is well worth to remember that

\beqn
[{\rm e}^{-{\bf i} H t}]_{\mathcal{A}_H}& = &
\left(
\begin{array}{cc}
j & 0 \\
0 & d \\
\end{array}
\right) \nonumber \\
j &=&
\left(
\begin{array}{cc}
{\rm e}^{-\widetilde{E}_1 {\bf i} t} & -{\bf i} {\rm e}^{-\widetilde{E}_1 {\bf i} t} t  \\
 0 & ~~{\rm e}^{-\widetilde{E}_1 {\bf i} t} \\
\end{array}
\right)
\nonumber \\
d & = &
\left(
\begin{array}{ccc}
 {\rm e}^{ -\widetilde{E}_3  i t} & 0 & 0 \\
 0 &{\rm e}^{-\widetilde{E}_4 i t} & 0 \\
 0 & 0 &{\rm e}^{-\widetilde{E}_5 i t} \\
\end{array}
\right),\nonumber \\
\label{evop}
\eeqn
so that, when writing $|I(t) \rangle$ of Eq.(\ref{init}), the coefficients $\widetilde{c}_\alpha(t)$ are given by

\beqn
\widetilde{c}_1(t) &=& {\rm e}^{- {\bf i} \widetilde{E}_1 t}\widetilde{c}_1  -{\bf i} {\rm e}^{- {\bf i} \widetilde{E}_1 t} t  \widetilde{c}_2,\nonumber \\
\widetilde{c}_k(t) &=& {\rm e}^{-{\bf i} \widetilde{E}_k t}\widetilde{c}_k, ~~~k=2,3,4,5.
\label{ct}
\eeqn

The deviation from the exponential$/$periodic behavior of Eq.(\ref{ct}) will be reflected on the time evolution of different physical observables. As an example, we can compute the survival probability, $p(t)$, of a given initial state as a function of time

\beqn
p(t) = | \langle I | I(t)\rangle |^2,
\label{prob}
\eeqn
where $| I(t)\rangle$ is defined in Eq.(\ref{init}).

In Figures 5 and 6, we present the results that we have obtained for the survival probability, $p(t)$ of Eq.(\ref{prob}), as a function of time. We have taken a system of $N=4$ particles. As initial state we have adopted a CSS, Eq.(\ref{istate}), with $(\theta_0,\phi_0)= (\pi/4,0)$ for Figure 5, and with $(\theta_0,\phi_0)= (\pi/8,0)$ for Figure 6. The time is scaled by $\kappa/\lambda$. In both Figures, the curves presented in Panels (a) have been calculated for $\kappa/\lambda=0.05$. For this value of the coupling ratio $\kappa/\lambda$, the spectrum of Hamiltonian of Eq.(\ref{hami}) is real, see Figure 4. This fact is reflected on the periodic behavior of $p(t)$. The curves displayed in Panels (b), have been calculated for $\kappa/\lambda= ~0.0739815$, this is the value at which the lower exceptional point takes place. Though, for this value of the ratio $\kappa/\lambda$, the spectrum is real (see Figure 4), the periodic pattern displayed by the curves of Panels (a) disappears due to behavior of the coefficients of Eq.(\ref{ct}). In Panels (c), we have plotted $p(t)$ for $\kappa/\lambda=0.1$. This value of $\kappa/\lambda$ is intermediate between the values at which exceptional points occur. The tendency to a stationary behavior is consequence of the appearance of complex eigenvalues in the spectrum. The curves of Panels (d) have been computed for the values of $\kappa/\lambda$ at which the second exceptional point takes place, $\kappa/\lambda=0.375$. We have taken $\kappa/\lambda=0.5$ for the curves drawn in Panels (e). At this value of $\kappa/\lambda$ the spectrum has one real eigenvalue, which dominates the behaviour of $p(t)$ for large values of $t$. Similar results, concerning the time evolution at exceptional points and not near them, have been presented in \cite{pe1,pe2}.

In Figure 7 and 8, we present the numerical results that we have obtained for the squeezing parameters and for the mean value of the components of spin, as a function of the time, at the exceptional points and away from them.
The Squeezing Parameters of Eq.(\ref{sqx}) in units of $[{\rm dB}]$, are presented in panels (a) and (b). The Mean Values of the components of the spin, $\langle S_k \rangle$, as a function of the time, are displayed in panels (c) and (d). We have considered an initial CSS with $(\theta_0,\phi_0)= (\pi/8,0)$.
In panels (a) and (c) of Figure 7, we show the results obtained $\kappa/\lambda=0.05$, and in panels (b) and (d) we plot the results obtained for the exceptional point $\kappa/\lambda=0.0739815$. For both values of $\kappa/\lambda$
the Hamiltonian has real eigenvalues. This fact is reflected in the oscillatory behaviour of the observables of the system. At the exceptional point  $\kappa/\lambda=0.0739815$, the periodic behaviour is lost, because of the structure of the coordinate $\widetilde{c}_1(t)$ of Eq.(\ref{ct}). In panels (a) and (c) of Figure 8, we show the results obtained for the exceptional point $\kappa/\lambda=0.375$, and in panels (b) and (d) we plot the results obtained for $\kappa/\lambda=0.5$. For both values of $\kappa/\lambda$
the Hamiltonian has complex conjugate pair eigenvalues. Thus, the system evolves to a squeezed steady state, which minimizes the corresponding uncertainty relations.

\begin{figure}[h!]
\includegraphics[width=6cm]{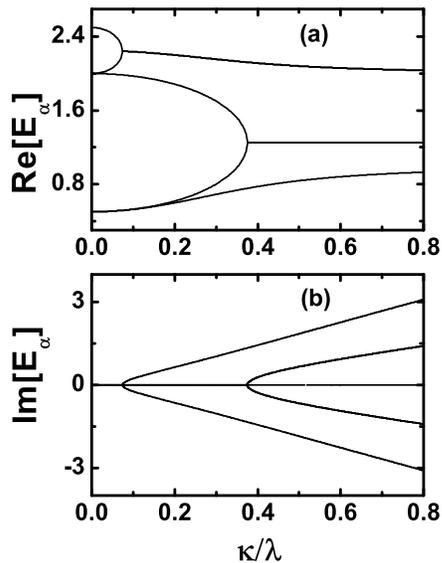}
\caption{Eigenvalues of the Hamiltonian of Eq.(\ref{hami}), for $N=4$ particles, as a function of the coupling ratio $\kappa/\lambda$. In panel (a) are shown the behaviour of the real part of the eigenvalues, while in panels (b) the imaginary part of each eigenvalue is presented. } \label{fig:fig4}
\end{figure}

\begin{figure}[h!]
\includegraphics[width=6cm]{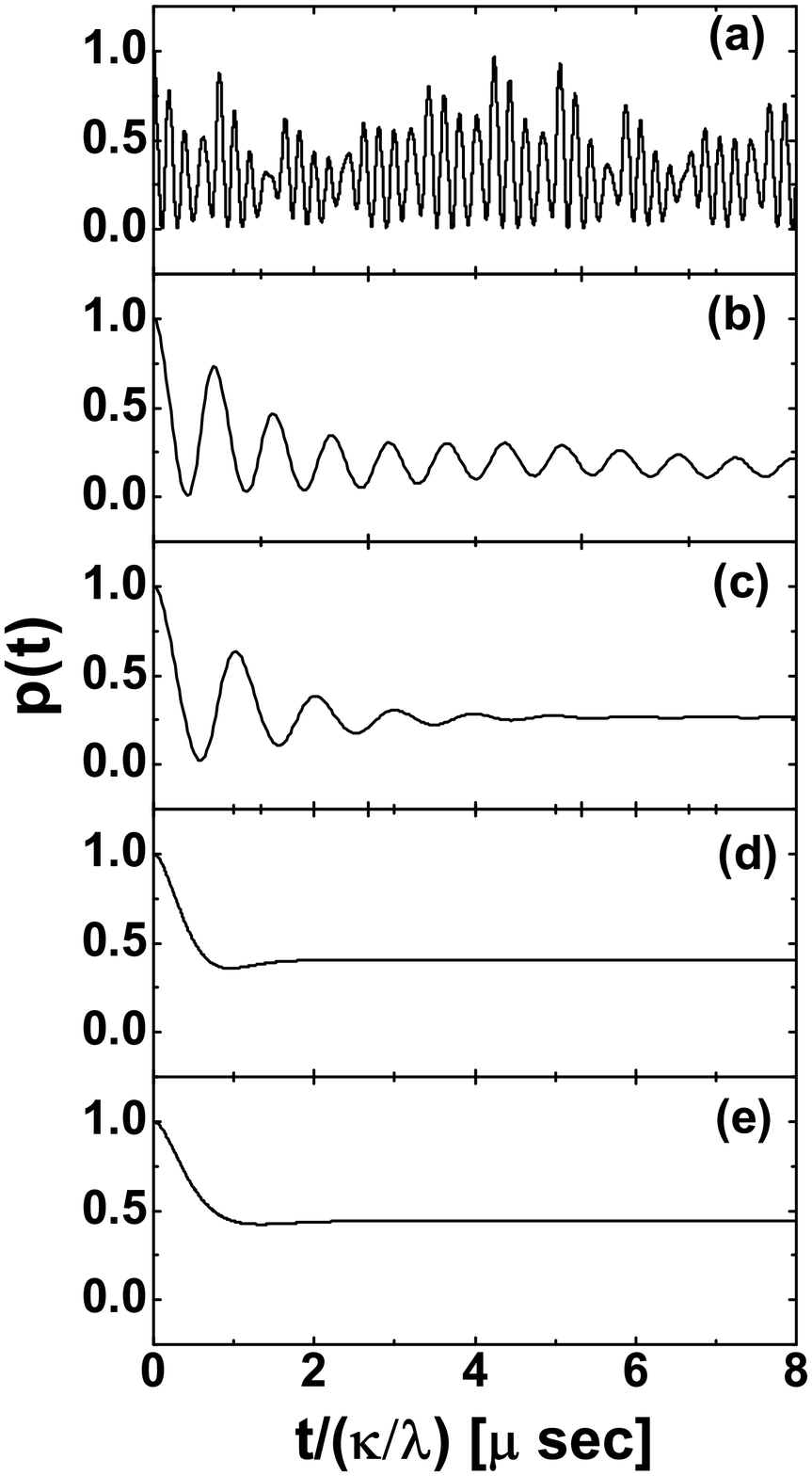}
\caption{Survival probability, $p(t)$ of Eq.(\ref{prob}), for a system of $N=4$ particles, as a function of time. The initial state is a CSS, Eq.(\ref{istate}), $(\theta_0,\phi_0)= (\pi/4,0)$. In panels (a)-(e), we show the results we have obtained for $\kappa/\lambda=0.05, ~0.0739815, ~0.1, ~0.373$ and $0.5$, respectively. } \label{fig:fig5}
\end{figure}

\begin{figure}[h!]
\includegraphics[width=6cm]{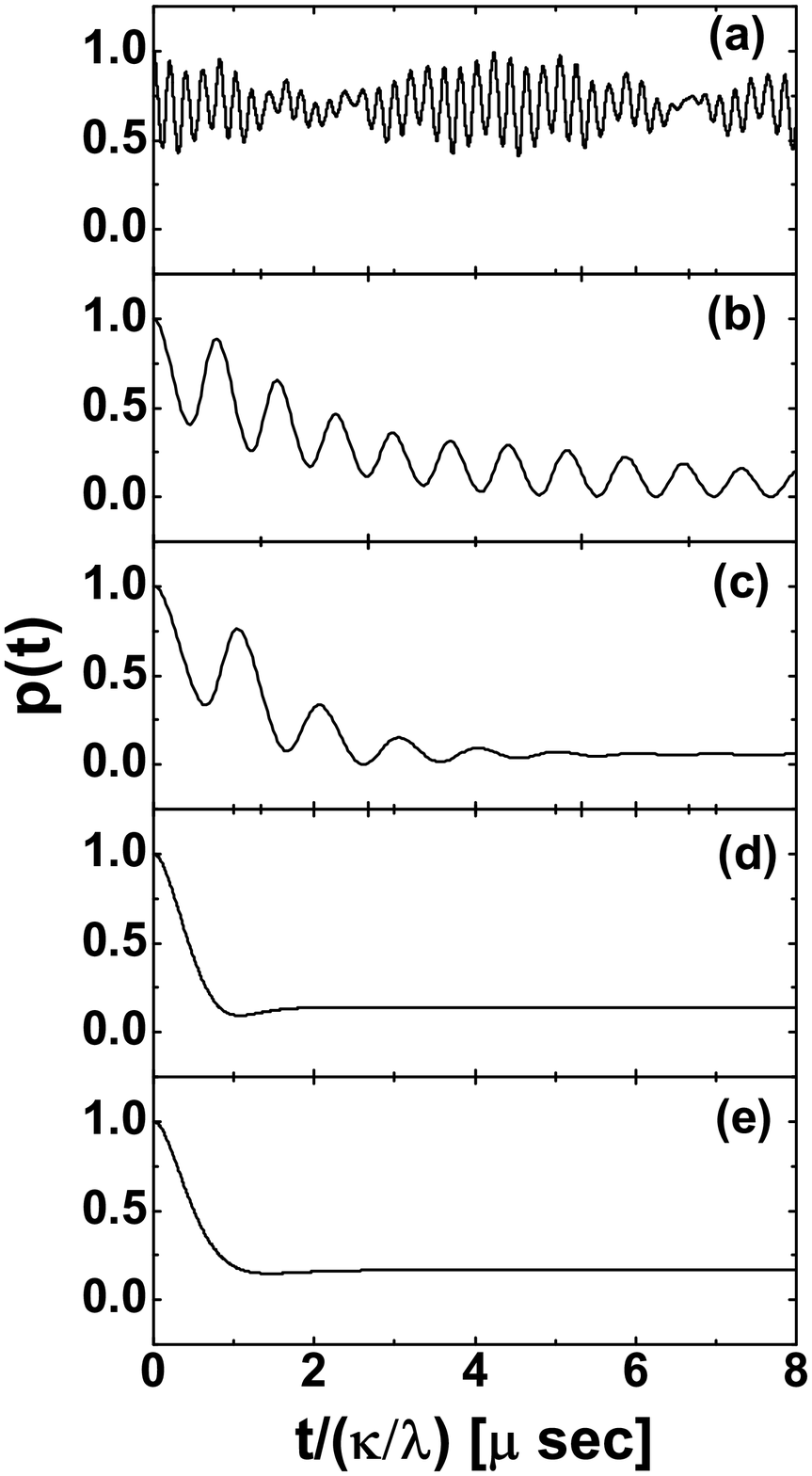}
\caption{Survival probability, $p(t)$ of Eq.(\ref{prob}), for a system of $N=4$ particles, as a function of time. The initial state is a CSS, Eq.(\ref{istate}), $(\theta_0,\phi_0)= (\pi/8,0)$. In panels (a)-(e), we show the results we have obtained for $\kappa/\lambda=0.05, ~0.0739815, ~0.1, ~0.373$ and $0.5$, respectively. } \label{fig:fig6}
\end{figure}

\begin{figure}[h!]
\includegraphics[width=6cm]{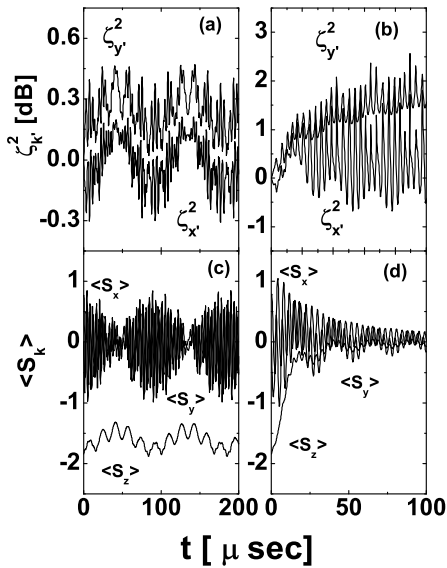}
\caption{Squeezing Parameters of Eq.(\ref{sqx}) as a function of the time, in units of $[{\rm dB}]$, are presented in panels (a) and (b). The Mean Values of the components of the spin, $\langle S_k \rangle$, as a function of the time, are displayed in In panels (c) and (d). We have considered a system with $N=4$ particles and an initial CSS with $(\theta_0,\phi_0)= (\pi/8,0)$.
In panels (a) and (c), we show the results obtained for $\kappa/\lambda=0.05$, while in
panels (b) and (d), we show the results obtained for the exceptional point $\kappa/\lambda=0.0739815$.} \label{fig:fig7}
\end{figure}

\begin{figure}[h!]
\includegraphics[width=6cm]{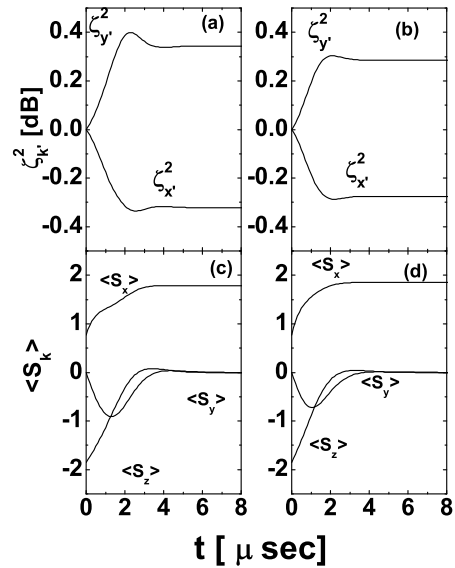}
\caption{
Squeezing Parameters of Eq.(\ref{sqx}) as a function of the time, in units of $[{\rm dB}]$, are presented in panels (a) and (b). The Mean Values of the components of the spin, $\langle S_k \rangle$, as a function of the time, are displayed in In panels (c) and (d). The parameters adopted are those of Figure 7. In panels (a) and (c), we show the results obtained for the exceptional point $\kappa/\lambda=0.375$, and in panels (b) and (d) we plot the results obtained for $\kappa/\lambda=0.5$, away from the exceptional point.} \label{fig:fig7}
\end{figure}

%\subsubsection{General non-hermitian Hamiltonian.}
As an example of a general non-hermitian Hamiltonian, let us consider another generalization of the OAT Hamiltonian \cite{disi,disib,nosap17}. That is

\beqn
H_S & = & H_{sp}+H_{OAT}+ H_L, \label{hami2} \\
H_{sp} &=& (\epsilon-{\bf i} \gamma)~ S_z, \nonumber \\
H_{OAT} & = &  \frac 1 2 \chi ~ S_z^2, \nonumber \\
H_{L}& = &  + ~ V ~ (S_x^2-S_y^2).  \nonumber
\eeqn

The Hamiltonian of Eq.(\ref{hami2}) consists of a
OAT term, $H_{OAT}$, plus a Lipkin-type, $H_L$, term. In addition, we shall assume that the particles of the system have a finite lifetime, which is given by the line-width $\gamma$ \cite{disi,disib,nosap17}.

From the physical point of view, the Hamiltonian $H_S$ of Eq.(\ref{hami2}) can be used to model a system of Nitrogen-Vacancy (NV) Centers in diamond \cite{nvint,nvint1,nvint2,marco,master1,nosap17}. An NV center has a ground state with spin $1$ and a
zero-field splitting D = 2.88 GHz between the $|1,0>$ and
$|1,\pm 1>$ states. If an external magnetic field,
along the crystalline axis of the NV center, is applied an additional
Zeeman splitting between $|1, \pm 1>$ sub-levels occurs. Then, it is possible to isolate the subsystem form by $|1,0>$ and $|1,1>$, so that the NV center can be modeled by a two-level system \cite{marco,master1}, through an effective spin-spin interaction \cite{nvint,nvint1,nvint2} of the form given in (\ref{hami2}).

From a mathematical perspective, if $\epsilon \in \mathds{R}$ and $\epsilon \ne 0$, the Hamiltonian of Eq.(\ref{hami2}) is not invariant under ${\mathcal PT}$ symmetry.
Observe that ${\mathcal P~ T}~(\epsilon~S_z)~ {\mathcal P}^{-1} {\mathcal T}^{-1}=-\epsilon ~S_z$.
Consequently, it has not complex-conjugate pair eigenvalues.

In Figure 9, we plot both the squeezing parameters, in units of $[{\rm dB}]$, and the mean value of the components of spin, as a function of the time. As an example, we have fixed the number of particles to $N=15$, and the value of the constants to $\epsilon=1.0,~\chi=2.88, ~V=0.26$ and $\gamma=0.02$, in units of [MHz].
Squeezing Parameters of Eq.(\ref{sqx}) as a function of the time, in units of $[{\rm dB}]$, are presented in panels (a) and (c). The Mean Values of the components of the spin, $\langle S_k \rangle$, as a function of the time, are displayed in In panels (b) and (d).  In panels (a) and (b), we show the results obtained for an initial CSS with $(\theta_0,\phi_0)= (\pi/4,0)$. For panels (c) and (d) we have taken $(\theta_0,\phi_0)= (\pi/8,0)$
Also, the initial coherent state evolves in time to a steady squeezed state.

\begin{figure}[h!]
\includegraphics[width=6cm]{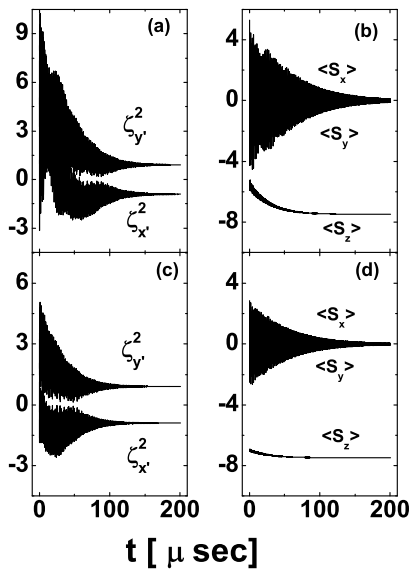}
\caption{Squeezing Parameters of Eq.(\ref{sqx}) as a function of the time, in units of $[{\rm dB}]$, are presented in panels (a) and (c). The Mean Values of the components of the spin, $\langle S_k \rangle$, as a function of the time, are displayed in In panels (b) and (d). We have fixed $N=15$, and $\epsilon=1.0,~\chi=2.88, ~V=0.26$ and $\gamma=0.02$ in units of [Mhz]. In panels (a) and (b), we show the results obtained for an initial CSS with $(\theta_0,\phi_0)= (\pi/4,0)$. For panels (c) and (d) we have taken $(\theta_0,\phi_0)= (\pi/8,0)$ .} \label{fig:fig8}
\end{figure}

The numerical results that we have presented suggest that we can enhance the effect of coherence of the physical system by adopting coupling constants in regime of real spectrum. On the other hand, if we want to achieve a steady squeezed state, we have to fix the parameters of the model in the region of complex spectrum. It should be noticed, as has been pointed in \cite{savannah}, that at exceptional points, due to dependance in time of the evolution operator, Eq.(\ref{evop}), deviations from the exponential decay are present.

 The generalization of the previous formalism to study infinite dimensional systems is not straightforward \cite{Bagarello book,bookazizov}. There are families of non-hermitian Hamiltonians for which their eigenfunctions and spectrum can not be used to complete the information of $H$ in the whole Hilbert space $\mathcal{H}$. The treatment of these problems involves another tools as generalized Riesz systems \cite{Inoue2016,bag2018,benkuz},  pseudospectrum \cite{Bagarello book}, unbounded metric operators and spectral functions for definitizable operators in Krein spaces \cite{spiegel,bag2018,mos2013,Langer04}. Also, the domain of the spectral functions is a non trivial issue to address \cite{carten}.

\section{Conclusions}\label{conclusions}

In this work, we have studied the time evolution of finite dimensional non-hermitian Hamiltonians. In doing so, we have constructed metric operators and the corresponding inner products.
In the case of pseudo-hermitian Hamiltonians, we have analyzed the regime of real and of complex-conjugate pair spectrum. Also, we have studied the time evolution of pseudo-hermitian Hamiltonians at exceptional points. We have made use  of the formalism of Krein Spaces to define inner products when dealing with pseudo-hermitian Hamiltonians with complex spectrum. As an example, we have studied the stationary behavior of  non-hermitian One Axis Twisting Hamiltonians. We have discussed the effect of decoherence in the different coupling schemes. As it is expected, we observe that the results depend drastically on the characteristic of the spectrum of the Hamiltonian. If the spectrum of the Hamiltonian is real, the observables of the system show a series of revivals as function of time. On the other hand, if the spectrum of the Hamiltonian includes complex eigenvalues, due to the effect of decoherence, the system evolves into a pointer state. We observe that at exceptional points deviations from the exponential decay form are present due to dependance in time of the evolution operator.
Work is in progress concerning the extension of the formalism to physical systems described by definitizable Hamiltonian operators in Krein spaces.

\begin{acknowledgments}
This work was partially supported by the National Research Council
of Argentine (PIP 282, CONICET) and by the Agencia Nacional de
Promocion Cientifica (PICT 001103, ANPCYT) of Argentina.
\end{acknowledgments}

\end{document}